 \documentclass[reprint, amsmath, amssymb, aps, prb]{revtex4-1}

\usepackage[]{graphicx}      % Include figure files
\usepackage{bm}             	% bold math
\usepackage{times}			% Times font
\usepackage{tabularx}
\usepackage{color}
\usepackage{dcolumn}		% Align table columns on decimal point
\usepackage{amsmath}
\usepackage{amssymb}
\usepackage{amsfonts}
\usepackage{times}
\usepackage{tabularx}
\usepackage{xcolor,colortbl}

\begin{document}
\title{Evolution of perpendicular magnetized tunnel junctions upon annealing}
\author{Thibaut Devolder}
\email{thibaut.devolder@u-psud.fr}
\affiliation{Institut d'Electronique Fondamentale, CNRS, Univ. Paris-Sud, Universit\'e Paris-Saclay, 91405 Orsay, France}
\author{S. Couet}
\author{J. Swerts}
\author{A. Furnemont}
\affiliation{imec, Kapeldreef 75, B-3001 Leuven, Belgium}

\date{\today}                                           
%%%%%%%%%%%%%%%%%%%%%%%%%%%%%%%%%%%%%%%%
%
%       Abstract
%
%%%%%%%%%%%%%%%%%%%%%%%%%%%%%%%%%%%%%%%%
\begin{abstract}
We study the evolution of perpendicularly magnetized tunnel junctions under 300 to 400 $^{\circ}$C annealing. The hysteresis loops do not evolve much during annealing and they are not informative of the underlying structural evolutions. These evolutions are better revealed by the frequencies of the ferromagnetic resonance eigenmodes of the tunnel junction. Their modeling provides the exchange couplings and the layers' anisotropies within the stack which can serve as a diagnosis of the tunnel junction state after each annealing step. The anisotropies of the two  CoFeB-based parts and the two Co/Pt-based parts of the tunnel junction decay at different rates during annealing. The ferromagnet exchange coupling through the texture-breaking Ta layer fails above 375$^{\circ}$C. The Ru spacer meant to promote a synthetic antiferromagnet behavior is also insufficiently robust to annealing. Based on these evolutions we propose optimization routes for the next generation tunnel junctions.
\end{abstract}

\maketitle

%%%%%%%%%%%%%%%%%%%%%%%%%%%%%%%%%%%%%%%%
%
%                Paper
%
%%%%%%%%%%%%%%%%%%%%%%%%%%%%%%%%%%%%%%%%

%%%%%%%%%%%%%%%%%%%%%%%%%%%%%%%%%%
%\section{Introduction}
Spin-transfer-torque magnetic random access memories (STT-MRAM) are considered to have the potential to become a unified non-volatile memory for embedded applications as it simultaneously can serve as a working memory and a storage memory for code and data \cite{kent_new_2015}. This technology is based on perpendicularly magnetized magnetic tunnel junctions (MTJ)  that are grown at the back end of line of CMOS and must thus sustain the associated 400 $^{\circ}$C final anneal. In addition to the free layer that stores the information, state-of-the-art MTJs \cite{gajek_spin_2012} rely on composite fixed systems that comprise many layers. A generic fixed system comprises three parts: a FeCoB-based layer in contact with MgO to ensure optimal spin transport properties, a high anisotropy multilayer that pins it through a coupling layer, and another hard multilayer coupled antiferromagnetically. The two hard multilayers are constructed in a synthetic ferrimagnet configuration to avoid the generation of stray fields that would destabilize the free layer and lead to asymmetric switching properties \cite{bandiera_comparison_2010, gopman_bimodal_2014}. Each magnetic layer and each coupling layer requires a careful optimization \cite{swerts_beol_2015} and must keep its properties upon annealing which is a challenging objective upon active investigation \cite{yakushiji_ultrathin_2010, gottwald_ultra-thin_2013, fang_tunnel_2015, gottwald_scalable_2015}. Unfortunately, the properties of the individual layers cannot easily be isolated and determined to guide further material optimization. In conventional magnetometry methods the information on the free and fixed systems is intertwined. Most often the MTJ functionality can only be measured at the device level with potential artefacts from patterning damages. When the junction in inoperative, it is often difficult to backtrace to the problem.

In this paper, we study the annealing evolution of each part of a state of the art MTJ based on Co/Pt multilayers, which are thought to be more temperature resistant than Co/Pd \cite{yamane_changes_1993} or Co/Ni based systems \cite{kurt_enhanced_2010}. We use vector network analyzer ferromagnetic resonance (FMR) to measure the stack eigenexcitations up to 2.5 T and 70 GHz and model them to deduce the anisotropies and the interlayer coupling of the essential layers of the MTJ at various stages of annealing. The identification of the most critical layers can be used to define optimization routes for the next generation magnetic tunnel junctions.

\begin{figure*}
\includegraphics[width=17 cm]{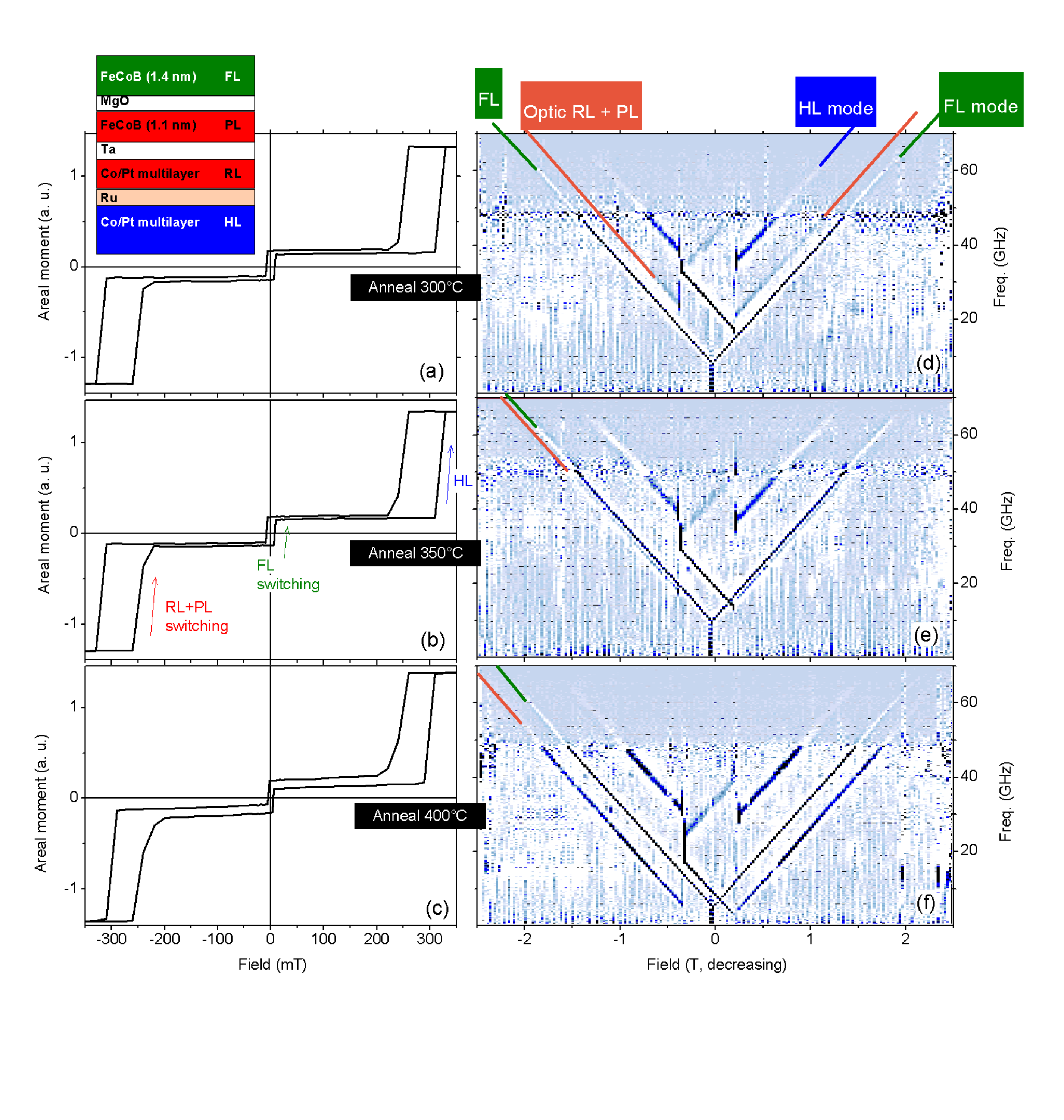}
\caption{Properties of the MTJ after 300, 350 and 400$^{\circ}$C annealing. (a), (b) and (c): easy axis loops. (d), (e) and (f): imaginary part of the magnetic permeability (color, arbitrary scale) versus out-of-plane field and frequency during a downwards field sweep. The change of contrast above 48 GHz is an experimental artefact. The mode labels refer to the repartition of the precession amplitude among the PL, RL and HL as modeled in Fig.~\ref{Fit300C}.  Inset in(a): sketch of the nominal stack.}
\label{Exp300350400C}
\end{figure*}

%%%%%%%%%%%%%%%%%%%%%%%%%%%%%%%%%%
%\section{Samples} \label{FilmProp}
Our objective is to understand the evolution of the fixed layers of perpendicularly magnetization MTJ upon high temperature annealing. For this, we use bottom-pinned MTJs (Fig.~\ref{Exp300350400C}(a), inset) of the following configuration \cite{devolder_time-resolved_2016}: seed / Hard Layer / Ru / Reference Layer / Ta  / Spin Polarizing Layer / MgO (RA=6.5~$\Omega.\mu \textrm{m}^2$, TMR =150 \%)/ Free Layer / cap, with compositions described in Table I. We study the fixed layers in state-of-the-art MTJÕs, i.e. including a free layer. This is meant to reflect the real environment of the fixed layers; indeed the properties of the fixed layer system (crystallization, thermal robustness...) can be influenced by the free layer on top. Our fixed system is constructed in a synthetic ferrimagnet configuration as commonly practiced for stray field compensation. It consists of three parts (Fig.~\ref{Exp300350400C}(a), inset): the 1.1 nm thick FeCoB spin polarizing layer (PL) whose \textit{fixed} character is ensured by a ferromagnetic coupling with the Co/Pt based reference layer (RL) though a Ta spacer. The RL is hardened by an antiferromagnetic coupling with a thicker Co-Pt based hard layer (HL) through a Ru spacer.  All samples have been annealed at 300$^{\circ}$C for 30 minutes in a field of 1 Tesla, followed by a rapid thermal annealing of 10 minutes at $T_a=$ 350, 375 or 400$^{\circ}$C.

%%%%%%%%%%%%%%%%%%%%%%%%%%%%%%%%%%

%%%%%%%%%%%%%%%%%%%%%%%%%%%%%%%%%%
%\section{Results}
Below and at $T_a = 350^\circ$C, the easy axis loops [Fig.~\ref{Exp300350400C}(a-c)] confirm that all layers have perpendicular magnetization. The PL and the RL are rigidly coupled and they switch in synchrony, which relates to the ferromagnetic interlayer exchange coupling $J_\textrm{Ta}>0$ though Ta. The negative sign of their coercivity recalls their antiferromagnetic coupling $J_\textrm{Ru} <0$ with the HL. These two features are maintained upon annealing at the first two temperatures, until a partial decoupling of the RL and the PL appears. The loop after 300 and 350 $^{\circ}$C are almost indiscernible [Fig.~\ref{Exp300350400C}(a-b)] and possess perfectly horizontal plateau when the \{RL+PL\} ensemble is antiparallel to the HL, disregarding the state of the FL). After the toughest annealing, there is a faint reduction of the HL coercivity. While this could originate either from a decrease of the HL anisotropy or from a decrease of the exchange through Ru we will see later that it results from a combination of the two effects. In addition, a rounding of the coercivity of the \{PL + RL\} ensemble is observed, and a slope and an opening appear at low fields. This is indicative of some reduction of the anisotropies within the \{PL + RL\} ensemble and indicates that these two layers do not switch in synchrony any longer. Transport properties deduced from current-in-plane-tunneling experiments at~$\pm 40 ~\mathrm{mT}$ indicate that the tunnel magnetoresistance is optimal after annealing at 350 and 375$^\circ$C. We could not determine to what extent the decrease of TMR at higher annealing was due to a degradation of the spin-transport properties or to a partial loss of perpendicular anisotropy within the PL. Annealing renders the oxide barrier more resistive (Table I).

All in all, the hysteresis loops show limited evolutions upon annealing and the tiny changes are hard to interpret.  This does not mean that the MTJ do not evolve upon annealing; indeed the coercivities are usually extrinsic and cannot be used in general to assess the values of the intrinsic properties like anisotropy and interlayer exchange. We will see hereafter that the stack properties change substantially even if the loops do not reflect it. The intrinsic properties can be inferred from the MTJ's ferromagnetic resonance eigenexcitations \cite{devolder_performance_2013}.  The dynamical magnetization properties were determined by vector network analyzer ferromagnetic resonance (VNA-FMR \cite{bilzer_vector_2007}) in the open-circuit total reflection configuration \cite{bilzer_open-circuit_2008}. The susceptibility spectra Fig.~\ref{Exp300350400C}(d-f)]  were recorded for fields up to 2.5 T and frequencies in the 1-70 GHz interval with post-treatments described in ref. \onlinecite{devolder_damping_2013}. 
 
The MTJ comprises four parts and thus it should yield four resonances. Unfortunately we could only detect three modes. The V-shaped modes [Fig.~\ref{Exp300350400C}(d-f), green labels] with narrow linewidths bend at the free layer coercivity and must thus be assigned to the FL. The FL properties can be deduced following a straightforward analysis as conducted in many places \cite{devolder_damping_2013}. The other modes are affected by the exchange through Ru and Ta such that they involve magnetization motion in all three parts of the fixed system. For reasons that will become clear upon modeling, we refer the highest frequency mode as HL mode (blue labels) and to the lowest frequency mode as the optical \{RL+PL\} mode (orange labels). \\
Upon annealing, the optical \{RL+PL\} mode decreases by typically 5 GHz per annealing step in both the parallel (P) and antiparallel (AP) configurations. The "HL" mode is first unaffected then decreases by 2 GHz at 375$^\circ$C (not shown) in both high field (P) and low field (AP) states. It then collapses by -19 GHz at the last annealing step in the AP state, and by -5 GHz in P. This pronounced difference between the P and AP configurations is indicative of a dramatic decrease of the antiferromagnetic coupling through Ru. None of the detected modes softens at the loops' characteristic fields. While this is expected at the coercivity of the HL which is anticipated to be extrinsic, it is surprising at the coercivity of the \{PL + RL\} ensemble where the antiferromagnetic coupling generally promotes a more reversible behavior \cite{devolder_performance_2013}. 

%%%%%%%%%%%%%%%%%%%%%%%%%%%%
\begin{figure}
\includegraphics[width=9 cm]{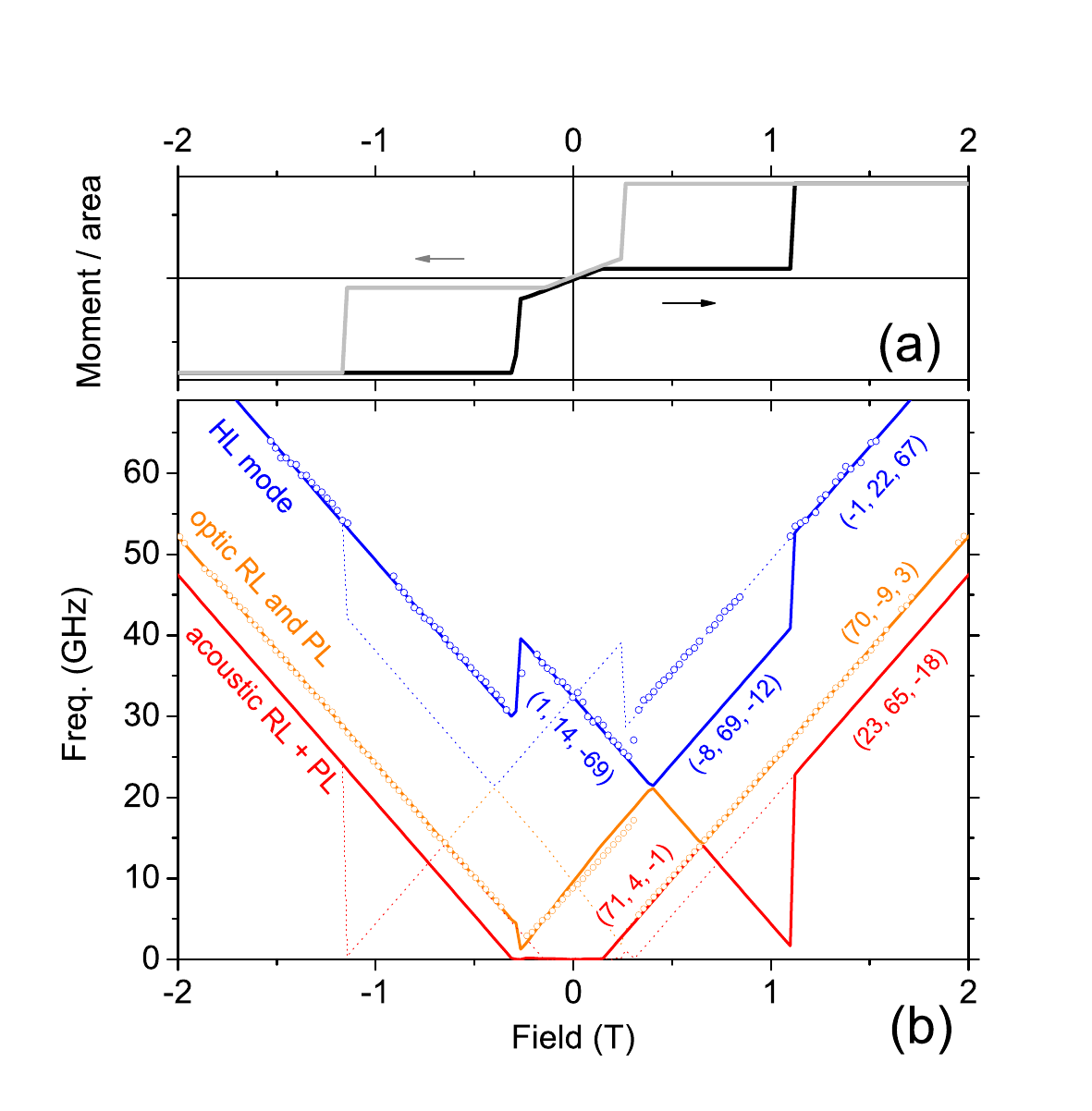}
\caption{Modeled properties of fixed part of the MTJ after 400 $^{\circ}$C annealing. (a) : Simulated hysteresis loop of the \{PL, RL, HL\} ensemble. (b) : Eigenmode frequencies. The circles are the experimental data extracted for increasing fields. The bold and dashed lines are the modeled frequencies predicted at increasing and decreasing field scans. The vectors denote the relative precession amplitude within the PL (first component), the RL and the HL (last component) in the collinear states.}
\label{Fit300C}
\end{figure}

\begin{table*}
  \centering
  \begin{tabular}{c|c|ccccc}
  \hline
Layer  & free layer & polarizing layer (PL) & & reference layers (RL)  &  & hard layers (HL) 	\\ 
Composition  & Fe$_{60}$Co$_{20}$B$_{20}$ & Fe$_{60}$Co$_{20}$B$_{20}$ & Ta &  [Co(5)/Pt(3)]$_{\times 4}$ /Co(5) & Ru &[Co(5)/Pt(3)]$_{\times 6}$ /Co(5)\\ 
Thickness ($t$, \r{A}) & 14 & 11 &  4 &   37 & 8.5 & 53 \\ \hline \hline
$M_S$ (A/m)  & $\dagger$ & $\dagger$ & & $\dagger$ & & $\dagger$\\   
300$^{\circ}$C &  $1.1\times 10^6 $ & $ 1.1\times10^{6} $ & & $0.76\times10^{6} $  & & $0.73\times10^{6}$ \\  
350$^{\circ}$C &  $id.$  & $ id. $ & & $ id. $  & & { $0.71\times10^{6}$ } \\ 
375$^{\circ}$C &  $id. $ & $ id. $ & & $ id. $  & & {$0.65 \times10^{6}$}  \\
400$^{\circ}$C &  $id.$ & $ id. $ & & $ id. $  & & {$0.63\times10^{6}$ } \\ \hline \hline
 
$ H_{k}$ (A/m)   &    &   & &   & & \\  
$\mu_0(H_{k}-M_S)$ (mT)   &  $\pm$ 5 mT  &  $\pm$ 25 mT & &  $\pm$ 25 mT  &   & $\pm$ 10 mT\\  \hline 
 
300$^{\circ}$C   & $ 1.34\times 10^6$  & $1.23\times 10^6$ & & $0.95\times 10^6$  & & $1.71\times 10^6$ \\  
    &   298 mT & 164 mT & &  238  mT  & & 1.23 T\\  \hline 

 350$^{\circ}$C   &  $1.39\times 10^6$ & $1.02\times 10^6$ & & $0.9\times 10^6$  & & $1.71\times 10^6$ \\   
    &  361 mT & -100 mT & & 176  mT & & 1.26 T\\  \hline 
  
 375$^{\circ}$C   & $1.34\times 10^6$  & $1.00\times 10^6$ & & $0.85\times 10^6$  & & $1.6\times 10^6$ \\    
  &  300 mT & -125 mT & & 113  mT& & 1.19 T\\  \hline 
 
400$^{\circ}$C &  $1.24\times 10^6$ & $0.94\times 10^6$ & & $ 0.82\times 10^6$  & & $ 1.38 \times 10^6$ \\   
   &  175 mT & -200 mT & & 75 mT & &  0.94 T\\  \hline \hline

$J$ (mJ/m$^2$) $\dagger$ & -  & & $J_{\textrm{Ta}}(\pm$ 0.03) & &$J_{\textrm{Ru}}(\pm$ 0.03) \\  
300$^{\circ}$C & -  & & 0.37 & & -1.32  \\  
350$^{\circ}$C & -  & & 0.38 & & -1.2  \\   
375$^{\circ}$C & -  & & 0.19 & & -1.1  \\   
400$^{\circ}$C & -  & & 0.07 & & -0.9 \\  \hline \hline
Transport properties & TMR  (\%) &  RA ($\Omega.\mu\textrm{m}^2$) & &  \\  
 300$^{\circ}$C &150 & 6.5 &  \\  
 350$^{\circ}$C & 175 & 6.7 &  \\   
 375$^{\circ}$C & 175 & 7.1 &  \\   
 400$^{\circ}$C & 156 & 8 &  \\  \hline 

    \end{tabular}
  \caption{Set of properties consistent with the eigenmode frequencies of the MTJ and measured transport properties. The symbol $\dagger$ recalls that the corresponding quantity was calculated assuming that the effective magnetic thicknesses is equal to the nominal thickness and stays constant during annealing.}
  \label{bilan}
\end{table*}

%\section{Model}
To get the quantitative properties of each part of the MTJ, we have fitted the observed mode frequencies to that of coupled macrospins. The ground state and the eigenexcitations of the fixed system are found in a manner similar to that of ref. \onlinecite{helmer_quantized_2010}, i.e. by  staying in the energy minimum during a field sweep and then by linearizing the LLG equations about the energy minima to infer the eigenexcitations of the system.  
We will assume that these modes arise from the 3 layers (PL, RL and HL) with their unknown anisotropies. The layers are coupled through interlayer exchange coupling factors $J$ through Ta and Ru (Table I) that act as two unknown coupling fields of the form $J / (M_s t)$. Apart from that of the hard layer, the magnetizations were considered constant during annealing since the other properties seem to evolve much more. 

We have 5 unknowns (the PL, RL and HL anisotropies and two coupling fields) and 5 relevant data (two frequencies in the parallel (high field) state, two others in the antiparallel state and one coercivity). We indeed constraint the fits to get the correct coercivity for the \{PL + RL\} ensemble.  For a fast and intuitive convergence, we use the precession eigenvectors to identify the parameters that influence the most a given mode. Let us illustrate this on the optic mode at high fields (Fig.~\ref{Fit300C}, orange curve). The relative amplitudes of precession are 70 in the PL, -9 in the RL and 3 in the HL. As a result, its frequency is thus much more influenced by the anisotropies of the PL and the RL than from that of the HL which hosts only a very tiny precession amplitude. Correlatively, the frequency of this mode is much more influenced by $J_\textrm{Ta}$ than by $J_\textrm{Ru}$. Since the PL and RL precess with phase opposition (optical excitation), we name this mode "optic \{RL+PL\}". In contrast, the mode named "acoustic" (Fig.~\ref{Fit300C}, red curve) is named so because it implies precessions that are well distributed and in-phase in the \{RL+PL\} ensemble. This mode is not detected experimentally except indirectly by the switching that it triggers when softening. Finally, the highest frequency mode (blue curve) is mainly hosted by the hard layer, hence its name.

The above approach was used to determine each layer's properties and their interlayer exchange coupling factors (Table I). In line with the conclusions of other studies \cite{lee_giant_2006, meng_annealing_2011, wang_rapid_2011, gan_origin_2011, miyakawa_impact_2013}, the anisotropy of the free layer is first improved, until it degrades above 375$^\circ$C. The FeCoB spin polarizing layer has the same nominal composition but its anisotropy degrades much earlier. This is probably linked to the lower crystalline quality of the spin polarizing layer that results from the fcc to bcc texture transition in the underlying Ta spacer. Noteworthy, the spin polarizing layer would be in-plane magnetized at soon as $T_\textrm{a} \geq 325^\circ$C without the coupling with the reference layer through Ta, as found already in similar systems \cite{le_goff_optimization_2015} .\\
The coupling through Ta is moderate and it decays substantially above 350$^\circ$C to become too weak to hold together the RL and the PL. Regarding the thermal budget, the Ta spacer is thus a weak point in the stack. 
The anisotropy of the Co/Pt  hard layer of the MTJ degrades for annealing above 375$^\circ$C but the magnetizations stays perpendicular for all annealing conditions. At first the decay of the anisotropy is compensated by a joint decrease of the magnetization, such that only the coupling field though Ru can unravel these two evolutions. 
 The joint decrease of magnetization and anisotropy is probably indicative of a gradual intermixing of the Co and Pt layers, which decreases the abruptness of the interfaces and consequently the interface anisotropy within the Co/Pt-based hard layer \cite{devolder_x-ray_2001}. 
The anisotropy of the Co/Pt  reference layer of the MTJ  is substantially weaker than that of the hard layers. This does not necessarily indicate a worse texture when grown on the Ru spacer than when grown on optimized seed layers; indeed our experience is that the main determinant to achieve high anisotropy is the number of repeats within Co/Pt multilayers. The anisotropy field of the RL also degrades upon annealing. 
Finally, the interlayer exchange coupling through Ru is initially very strong but it decays regularly.  It thus constitutes also a part of the stack that needed improvement to meet the constraints of the  thermal budget. This decay is consistent with older studies \cite{lee_giant_2006} done on in-plane magnetized systems.

%\section{Summary and concluding remarks}
In summary, we have studied the evolution of perpendicularly magnetized tunnel junctions upon annealing up to 400 $^{\circ}$C. The hysteresis loops do not evolve much during annealing and cannot reveal the underlying structural evolutions. The ferromagnetic resonance eigenmodes indicate however substantial evolutions within the stack, with significant lowering of the anisotropies of both the Co/Pt-based parts and of the CoFeB-based parts of the tunnel junction. The exchange coupling through Ta is not robust to annealing at 400$^\circ$C, which argues for advanced texture breaking layers enriched with Co and Fe that seem to promote stronger ferromagnetic coupling\cite{gottwald_paramagnetic_2013, gottwald_scalable_2015} .  The antiferromagnetic exchange coupling through Ru is also decaying regularly. This diagnosis argues for the use of thinner Ru spacers close to the first antiferromagnetic exchange maximum at 0.48 nm,  \cite{zhao_perpendicular_2008} to start from a higher coupling. Another option is the insertion of Pt dopants around the Ru spacer \cite{bandiera_enhancement_2012} which could increase the anisotropies of the Co/Pt multilayers by suppressing the suboptimal contribution of the Co/Ru interfaces \cite{miyawaki_perpendicular_2009}. This would also provide extra resistance to annealing \cite{bandiera_enhancement_2012}  with a marginal decrease of interlayer exchange coupling. An additional research route would be to replace Ru spacers by Rh spacers that exhibit giant interlayer exchange coupling \cite{zoll_preserved_1997, blizak_interlayer_2015} but whose resistance to annealing is yet to be explored.

%%%%%%%%%%%%%%%%%%%%%%%%%%%%%%%%%%%%%%%%%%%%%%%%%%%%%%%%%%%%%%
%%%%%%%%%%%%%%%%%%%
%%%%%%%%%%%%%%%%%%%	
%\bibliography{bib.bib}


\end{document}